\documentclass{acm_proc_article-sp}

\newcommand{\ket}[1]{\left|#1\right\rangle}

\begin{document}

\title{An introduction to Fault-tolerant Quantum Computing\titlenote{(Does NOT produce the permission block, copyright information nor page numbering). For use with ACM\_PROC\_ARTICLE-SP.CLS. Supported by ACM.}}
%
%
%
%
%

\numberofauthors{2} 
%
\author{
%
%
\alignauthor
Alexandru Paler\\
       \affaddr{University of Passau}\\
       \affaddr{Innstr. 43}\\
       \affaddr{Passau, Germany, 94032}\\
       \email{alexandru.paler@uni-passau.de}
\alignauthor
Simon J. Devitt\\
       \affaddr{Ochanomizu University}\\
       \affaddr{2-1-1, Otsuka, Bunkyo-ku}\\
       \affaddr{Tokyo, 112-8610, Japan}\\
       \email{devitt1@mac.com}
}

\maketitle
\begin{abstract}
In this paper we provide a basic introduction of the core ideas and theories surrounding 
fault-tolerant quantum computation.  These concepts underly the theoretical framework of 
large-scale quantum computation and communications and are the driving force for many 
recent experimental efforts to construct small to medium sized arrays of controllable 
quantum bits.  We examine the basic principals of redundant quantum encoding, required 
to protect quantum bits from errors generated from both imprecise control and environmental 
interactions and then examine the principals of fault-tolerance from largely a classical framework.  
As quantum fault-tolerance essentially is avoiding the uncontrollable cascade of errors caused by 
the interaction of quantum-bits, these concepts can be directly mapped to quantum information.
\end{abstract}




\section{Introduction}
Fault-tolerant, error corrected, digital quantum computing underpins a significant worldwide effort 
to construct viable, commercial quantum computing systems \cite{LJLNMO10}.  The size of such error corrected machines is 
somewhat daunting for a field that has only managed to experimentally fabricate arrays of up to about ten functional 
quantum-bits (qubits) \cite{K15,DSMN13}.  However, the theoretical framework for fault-tolerant quantum computing has 
existed for nearly 20 years and is very well understood and quantum computing is competing with the 
vast classical computing power currently in existence.  It would be unreasonable to believe (even given the apparent 
computational power quantum information processing has over classical computing), that a small, error prone array 
of qubits could computationally outperform a classical system comprising of potentially millions of computing cores, each 
itself containing billions (or even trillions) of transistors.  

Fault-tolerant quantum computing refers to the framework of ideas that allow qubits to be protected from quantum errors 
introduced by poor control or environmental interactions (Quantum Error Correction, QEC) and the appropriate 
design of quantum circuits to implement both QEC and encoded logic operations in a way to avoid these errors 
cascading through quantum circuits \cite{nielsen2010quantum}.  By avoiding a cascade of errors, there becomes a point (when the fundamental accuracy 
of individual qubits is high enough), where QEC is correcting more errors than are being created.  Once this {\emph threshold} has been achieved, expanding the size of the protective quantum will exponentially decrease the failure of the encoded 
information and allows us to achieve arbitrarily long quantum algorithms implemented with noisy devices.  

In this paper we will provide a basic introduction to some of the key principals of QEC and then pivot into a discussion 
about fault-tolerance that have been investigated in the classical computing world.  As the goal of fault-tolerance is 
to prevent errors to cascade uncontrollably, a large amount of classical work can be easily transferred to the quantum 
world.  

\section{Quantum Computing}
In this section we present the basic mathematical framework for qubits, quantum gate operations and 
quantum circuits.  Further details can be found in a number of papers \cite{DMN13} and 
books \cite{nielsen2010quantum} both from 
the physics community and the computer science community.  

Quantum circuits represent and manipulate information in \emph{qubits}. A single qubit has an associated \emph{quantum state} $\ket{\psi}= (\alpha_0, \alpha_1)^T = \alpha_0\ket{0} + \alpha_1\ket{1}$. Here, $\ket0 = (1, 0)^T$ and $\ket1 = (0, 1)^T$ are quantum analogons of classical logic values 0 and 1, respectively. $\alpha_0$ and $\alpha_1$ are complex numbers called \emph{amplitudes} with $|\alpha_0|^2 +|\alpha_1|^2 = 1$.

\emph{Quantum measurement} is defined with respect to a basis and yields one of the basis vectors with a probability related to the amplitudes of the quantum state. Common measurements are known as $Z$- and $X$-measurements. $Z$-measurement is defined with respect to basis $(\ket0, \ket1)$. Applying a $Z$-measurement to a qubit in state $\ket{\psi}= \alpha_0\ket{0} + \alpha_1\ket{1}$ yields $\ket0$ with probability $|\alpha_0|^2$ and $\ket1$ with probability $|\alpha_1|^2$. Moreover, the state $\ket\psi$ \emph{collapses} into the measured state (i.e. only the components 
of $\ket{\psi}$ consistent with the measurement result remains). $X$-measurement is defined with respect to the basis $(\ket+, \ket-)$, where $\ket{\pm} = \frac{1}{\sqrt2}(\ket{0} \pm \ket{1})$.

A state may be modified by applying single-qubit \emph{quantum gates}. Each quantum gate corresponds to a complex unitary matrix, and gate function is given by multiplying that matrix with the quantum state. The application of $X$ gate to a state results in a \emph{bit flip}: $X(\alpha_0, \alpha_1)^T = (\alpha_1, \alpha_0)^T$. The application of the $Z$ gate results in a \emph{phase flip}: $Z(\alpha_0, \alpha_1)^T= (\alpha_0, -\alpha_1)^T$.  A $Y = iXZ$ gate can be thought of as both a bit-flip and a phase-flip 
together on an individual qubit.  These three gates are important in the context of quantum errors.

The exponentiation of the Pauli matrices results in the rotational gates $R_x$, $R_y$, $R_z$ parameterised by the angle of the rotation \cite{nielsen2010quantum}. Hence the bit flip is a rotation by $\pi$ around the $X$-axis, implying that $X=R_x(\pi)$, and the phase-flip is a rotation by $\pi$ around the $Z$-axis, such that $Z=R_z(\pi)$. The Hadamard gate is $H=R_z(\pi/2)R_x(\pi/2)R_z(\pi/2)$ 
and can be used to take a computational state $\{ket{0},\ket{1}\}$ into superposition states, $\{\ket{\pm} = (\ket{0}\pm \ket{1})/\sqrt{2}$, a state with no classical analogue

In the context of fault-tolerant quantum computation, only one interaction gate is needed; the controlled-not (CNOT) gate.  
The CNOT gate is the quantum analogue of a binary XOR operation and is designed to bit-flip the state of a target qubit, 
conditional on the state of a control qubit.  This gate can be employed on certain quantum states to prepare 
{\em entangled states}.  For example, CNOT$(\ket{0}+\ket{1})\ket{0}/\sqrt{2} = (\ket{00}+\ket{11})/\sqrt{2}$.  This state, 
known as a Bell state has the property that measuring one of the qubits produces a random result ($\ket{0}$ or $\ket{1}$ with 
a 50:50 probability), but once the state of one qubit is measured, the state of the other qubit is also determined.  

It is well known that the ability to perform arbitrary rotations around two orthogonal axis (e.g. $R_z(\theta_1)$ and 
$R_x(\theta_2)$ for arbitrary $\{\theta_1,\theta_2\}$) and to couple arbitrary pairs of qubits with a CNOT gate is 
sufficient to realise any $N$-qubit unitary operation.  This gate set is therefore quantum universal.

\section{Errors and Quantum Error Correcting Codes}
There are two important differences between classical error correction and quantum error correction.  The first is the 
no-cloning theorem \cite{Wz82}, which states that is is impossible to perfectly copy an unknown quantum state.  i.e. there 
is no operation that satisfies $U\ket{\psi}\ket{0} = \ket{\psi}\ket{\psi}$ for an unknown $\ket{\psi}$.  Therefore, we are 
unable to protect arbitrary quantum states against errors by simply making multiple copies.  Secondly, any measurement of 
an arbitrary quantum state will {\em collapse the wavefunction} describing the state.  Hence protecting errors in an encoded 
piece of quantum information by measuring a certain subset of the encoded block will irrevocably destroy the information 
content of that state.  Therefore we need a slightly new mechanism to protect encoded quantum information.  

The foundation of quantum error correction is still based on classical coding theory, however we need to design codes 
in a slightly different manner.  This is due both to the restrictions of what we can theoretically do with quantum information, but 
also due to the possible errors that can affect qubits.  Unlike classical bits, which can only experience a flip between 
$\ket{0} \leftrightarrow \ket{1}$, qubits can also experience phase errors $(\ket{0}+\ket{1})/\sqrt{2} \leftrightarrow 
(\ket{0}-\ket{1})/\sqrt{2}$.  Additionally, errors do not occur in a discrete manner.  They are most often continuous errors, 
such as a rotation around the $X$ axis by some angle $\epsilon$ or some incoherent (non-unitary) error caused by 
the interaction with the outside environment.    

Due to expedience we will only present the formalism for {\em coherent errors}, those that can be represented by a unitary gate.  
In this case, an error operator, $E$, acting on a qubit, $\ket{\psi}$ can be decomposed into a linear superposition of $X$ gates, $Z$ gates and both $Y=iXZ$, $E\ket{\psi} = a_1\ket{\psi} + a_2X\ket{\psi}+a_3Z\ket{\psi} + ia_4XZ\ket{\psi}$.  If we could 
{\em magically} measure {\em if} an $X$ and/or $Z$ error occurred on a qubits (via some type of quantum measurement), 
the state would collapse to the state $\{\ket{\psi}, X\ket{\psi}, Z\ket{\psi}, XZ\ket{\psi}\}$ with a probability of 
$\{|a_1|^2, |a_2|^2, |a_3|^2, |a_4|^2\}$.  This converts a possible continuous quantum error into a discrete $X$ 
and/or $Z$ gate.  While the errors themselves are continuous (for very small errors, $|a_1|^2 \approx 1$), this  
determination of what type of error 
has occurred converts small errors into discrete bit- or phase-errors with small amplitudes converted to small probabilities for 
such results to be observed.  The question is, how do we detect if some type of discrete error has occurred?  

This detection occurs through the idea of redundant encoding with two classical codes.  One is designed to detect 
$X$-errors and one is designed to detect $Z$-errors without having to, necessarily, decode the codespace.  Detecting an 
error indirectly for bit-flips is commonplace in classical computer science and was usurped for the quantum regime.  The 
simplest example is the bit-flip code, with basis states given by $\ket{0}_L = \ket{0}^{\otimes N}$ and $\ket{1} = 
\ket{1}^{\otimes N}$, where the $^{\otimes N}$ notation simply meaning $N$ copies of the qubit.  The basic idea is 
that given a given codified, $N$, the number of physical flips needed to turn $\ket{0}_L \leftrightarrow \ket{1}_L$ 
scales linearly with $N$.  Again, in the quantum regime, we are not allowed to directly measure {\em any} subset 
of qubits in the code block.  So we need a different method to identify errors.  In the context of the bit flip code, we notice 
a certain property, namely that for both basis states, pairwise bit-parity in the code block is even (i.e. calculating the 
parity of any two bits via modulo addition for the $\ket{0}_L$ and $\ket{1}_L$ state is even).  If such a comparison ever results 
in an odd value, we know an error has occurred without actually knowing if we started with the $\ket{0}_L$ or $\ket{1}_L$ state.  
This is what we need.  Therefore, we need a way to calculate the parity of any two qubits in the code block without directly 
measuring the qubits themselves.  We can do this via the circuit in Fig. \ref{fig:parity}.  This circuit introduces an ancilla qubit 
that is initialised, interacted with a pair of qubits in the code block and measured.  The result of the measurement on the ancilla 
(either $\ket{0}$ or $\ket{1}$) will determine the parity of the two qubits (odd or even), and also {\em force} 
these two qubits to be in an even or odd parity state if they were not beforehand.  

The principal of a codespace within quantum computation is to construct encoded codewords that are {\em always} in 
certain, well defined parity states regardless of the state of the encoded information.  Physical errors will then perturb 
encoded information away from these well defined parities which can be detected without determining any information 
regarding the encoding.  

Returning back to the example of a redundancy code, the two encoded states $\ket{0}_L$ and $\ket{1}_L$ are 
constructed to be even parity states of any pairwise $Z$ operators.  i.e. applying the operator $Z_iZ_j$ for any $i$,$j$ $\in N$ 
returns the same state, $Z_iZ_j\ket{0,1}_L = \ket{0,1}_L$.  Bit flip errors result in states which violate this condition.  For example, a bit-flip on qubit one of the encoded block will result in $Z_1Z_j\ket{0,1}_L = -\ket{0,1}_L$, $\forall j$.  Hence 
if we measure the parity of any of these operators and we find an odd result, we know that some type of error has occurred.  
Determining a {\em location} for the error and how many unique errors we can identify depends on the size of the code block, 
$N$.  The parity of pairwise checks of the $Z_iZ_{i+1}$ operator will allow us to uniquely locate individual errors
and the number of errors we can successfully correct scales linearly with the number of qubits in 
the code block.  For $N$ qubits, we are able to uniquely correct $(N-1)/2$ errors.  

This example illustrates how we handle bit-flip errors within quantum information, what about phase errors?  In 
quantum information, phase-flip errors work exactly the same way as bit-flip errors if we take our compuational 
states as $\ket{\pm} = (\ket{0}\pm \ket{1})/\sqrt{2}$. i.e. a $Z$-error will take $\ket{+}\leftrightarrow \ket{-}$.  Hence 
if we used a redundancy code of the form $\ket{0}_L = \ket{+}^{\otimes N}$ and $\ket{1}_L = \ket{-}^{\otimes N}$ 
and instead of checking the parity of the $Z_iZ_{i+1}$ operator, we check the parities of the $X_iX_{i+1}$ operator for 
$i,\in (N-1)$, then everything works exactly the same way.  Therefore, a redundancy code either using $\ket{0,1}$ or 
$\ket{+,-}$ states will allow us to either protect encoded information against $X$-errors or $Z$-errors.  A full quantum 
error correction code therefore combines two classical codes, independently responsible for bit-errors and 
phase-errors.  

The Shor code is the simplest example of this \cite{S95}.  In the Shor code we essentially have one redundancy code 
embedded within another.  The code encodes a single qubit of information into nine physical qubits.  The basis 
states are given by,
\begin{equation}
\begin{aligned}
\ket{0}_L &= \frac{1}{2\sqrt{2}}(\ket{000}+\ket{111})(\ket{000}+\ket{111})(\ket{000}+\ket{111})\\
\ket{1}_L &= \frac{1}{2\sqrt{2}}(\ket{000}-\ket{111})(\ket{000}-\ket{111})(\ket{000}-\ket{111})
\end{aligned}
\end{equation}

We have three blocks of three qubits that effectively act as a distance three redundancy code to correct bit flips 
in the way that we described above.  This allows us to correct a single bit flip error in any one of the three blocks.  
In principal, this code can correct for three bit-flip errors (provided each error occurs in a separate block), however 
in general, as two or more errors can occur in a single block the code is described as only having the capacity for 
correcting a single bit flip error.  Phase errors are corrected via the three blocks and comparing the parity of pairwise 
blocks.  Again, the code, in principal, can correct for more than one phase error (provided errors occur in specific 
locations), but in general only a single arbitrary error is deterministically correctable.

\section{Fault-Tolerance}
The effectiveness of QEC depends on how we implement quantum circuits to realise the correction code.  As with classical 
computing, interactions between qubits during the execution of the circuit leads to the {\em copying} of errors.  If this 
happens in an uncontrolled manner, the QEC code is overwhelmed and the computation will fail.  Therefore, we need to be very careful when designing circuits such that this does not occur. While there is not often a direct comparison between classical and quantum information processing, the principals of fault-tolerance in a classical framework is easily transferable.  We will 
therefore focus on some general principals of fault-tolerance in the classical world which can be mapped directly to quantum 
computing.  

\section{Short parallel to distributed systems}

The need for fault-tolerance in quantum computing can be introduced by drawing parallels with distributed systems. The following insights do not extend classical distributed systems to quantum ones \cite{van2014quantum}, but propose the problem of quantum fault-tolerance to be formulated without introducing quantum information and quantum computing. The discussion will focus on processes and communication over point-to-point links. 

The distributed system is modelled as crash-stop \cite{cachin2011introduction}, such that the processes can crash and never return to live, there is at least one fault detector module in the system and the communication links are perfect (messages are not lost, duplicated or inserted by fault). Additionally, the distributed system includes a fault corrector, which is informed by the fault detectors about faults requiring correction.

The main target of fault-tolerance, as presented in these sections, is to \emph{control the propagation of faults} between processes. The presentation will focus on describing the distributed system elements, and the way these interact, but without delving into details about the liveness and safety properties of the presented protocols.

\subsection{The processes}

A process holds an abstract object $q$ and a black-box that consists of a coin, \emph{two} Boolean values and a real value $\tau$. The black-box is a model for the process faults. The Boolean values $b$ and $p$ are the results of tossing the coin twice. The value $true$ indicates heads, and $false$ stands for tails. 

None of the processes is aware of 
the Boolean values inside the box. Furthermore, the processes do not control the black-box and the coin toss. The coin toss is randomly performed (with a probability $\tau$), meaning that none, a single or both values are generated at undetermined execution points of the process. Therefore, there may be none or multiple coin tosses during the lifetime of a process. In order to simplify the modelling, the probability $\tau$ is equal for all the processes, and process failures are not correlated.

The actual state of a process is computed by the function $q_a=q+b+p$, where the $+$ operation is the addition of the two possible random faults modelled as Boolean values
A \emph{correct} process is the one for which $q_a = q$, while, if $q_a \neq q$, the process is called \emph{faulty}. In the following, $q$ will refer both to the object and to its state, depending on the context it is used in. It should be noted that object $q$ is noncopyable\footnote{Similarly, in C++ a private copy constructor and copy assignment operator are required for such classes.}, meaning that once constructed, it can be either transformed locally or through distributed computations, but cannot be copied between processes.

During process initialisation it possible to initialise the black-box values and to set the state of $q$ to a state chosen from a discrete set $sq \in S$. Once the black-box was set up, there is no guarantee that the values had not been changed by the coin tosses.

A terminating process returns a state $so \in S$. During termination, the current $q$ is classified against all the $S$-states, and the closest $so$ is returned. The classification is probabilistic, meaning that if there is no state $so$ for which $so=q$, then it could happen that the returned state $so$ is actually \emph{orthogonal} to the state $so'$ that was classified as next to $q$.

Once a process is terminated, it cannot be brought back to live and, if one would like to approximate the probability of $q$ being in any of the states of $S$, the complete distributed computation has to be repeated. More flexible initialisation and termination procedures could benefit from a larger set $S$, but this will negatively impact the practicality of system implementation.

\subsection{Communication and operations}

The Boolean black-box values of a process can be  either read and communicated to other processes, or updated by a second process. A single process is not able to read its values and correct them. Value correction is performed only through coordinated communication during a procedure similar to consensus \cite{cachin2011introduction}.

The processes are naive: if one of their Boolean values is considered heads, then the same happens to the other processes. The processes are also lazy, meaning that the communication of heads-values has an associated cost, such that \emph{cost(heads)=1} and \emph{cost(tails)=0}.

Ideally, every process should have its Boolean values always set to $false$ (tails). For this reason, in general, the process avoids initialising their black-boxes with heads values. Processes are not byzantine, and each time one communicates, it will try, based on the process values, to convince its partener to either flip or keep one of its values. More specifically, a communication step between two processes $x$ and $y$ is performed in two rounds: firstly, $x$ sends its $b$ value to $y$ requesting it to update its $b$ value to $b_x \oplus b_y$. Afterwards, process $y$ sends its $p$ value to $x$ asking an update of its $v$ value to $v_x \oplus v_y$. The $\oplus$ function models the behaviour of a coin flip: the coin returns to its initial value after two flips. During the third communication round the control process applies the distributed abstract operation $e$: the update of the state $q_y$ (target) is a function of state $q_x$ (control). The $e$-operation will not be detailed in the context of the analogy with distributed systems, and its cost is considered zero.

Besides communicating and performing the distributed operation $e$, the process can perform local (intra-process) operations that transform the state of the local object $q$.

\subsection{Distributed processing}

As previously introduced, inter-process comunication is both an attempt to correct the black-box values and, at the same time, a distributed computation. The simplest distributed system executing a single communication step consists of two processes, which we will call the \emph{control} and the \emph{target}. The control initiates the communication, thus is the requester during the first communication round, and the target is the requester of the second communication round. Additionally, a process can be both control and target during separate communication steps.

Generally, a distributed algorithm represens a series of inter-process communication steps and local operations. There are at least two types of algorithms: 1) distributed correction, where processes communicate only to correct their Boolean values; 2) distributed computation, in which processes try to solve a computational problem. Algorithms compliant with the first option generally consist only of communication steps and no local operations. A well-defined correction protocol (see Section~\ref{sec:fd}) is the execution of coordinated communication. Distributed computations (the second option) include intra-process operations but neglect (do not coordinate) the effect of the two correction rounds in each communication step. The result is that uncoordinated correction can lead to \emph{propagation of faults}: the heads values are being transferred, without the processes having noticed, from a faulty process to a correct one (see Section~\ref{sec:trans}).

\subsection{Fault-tolerant processes}

In general, fault-tolerance is achieved by using an hierarchic (layered) approach. Assuming the failure probability $\tau$ of a process, a set of two processes will fail simultaneously with probability $\tau^2$. The introduction of redundancies is the key of achieving fault-tolerance, and there are two types of possible redundancies: 1) computational redundancy, where the same computation is repeated sequentially for multiple times; 2) resource redundancy, where multiple processes are abstracted as a single logical process and the component processes are executed in parallel.

Computational redundancy is the equivalent to executing a distributed algorithm in epochs, and to guaranteeing that after a certain number of epochs a property of the algorithm is achieved. Resource redundancies are generally used when at least $f$ faulty processes are needed to be tolerated. For example, the uniform epoch consensus algorithm from \cite{cachin2011introduction} requires $N$ processes with $N>2f$.

Majorities (quorums \cite{cachin2011introduction}) are the most common option for checking the introduced redundancies. The simple majority ($N/2 +1$) of $N$ objects (processes, bits etc.) is used to introduce the fault-tolerant quantum computing in the following: a fault-tolerant logical process is constructed from three (or more) component processes (called components), and the logical process is able to tolerate at most one faulty component. The computation of quorums is detailed in Section~\ref{sec:compred}.

Due to the fact that the $q$ objects of each process are noncopyable increases the difficulty of implementing fault-tolerance through redundancy.  Copying the $q$ state of an existing process to a newly initialised one is not possible. As a result, separate components are initialised into the same state $q \in S$ at the start of the distributed algorithm and exactly the same operations are applied on their objects.

This work presents, without loss of generality, how the construction of logical processes is performed using triple-modular redundancy (TMR) \cite{koren2010fault}. The logical state $q_l$ of a logical process is a sequence of $n$ (in this work $n=3$) component process states: $q_l=\prod_{i=0}^n q_i$. Transforming $q_l$ represents the transformation of each $q_i$.

The repetition code is the TMR counterpart in the field of error-checking and -correction methods. As a consequence, the logical state $q_l$ should be interpreted as the encoding of one of the component process states (the states of the components are equivalent). The repetition code can be replaced with more powerful codes like the Hamming code or surface codes \cite{fowler2012surface}, but this aspect is not further addressed in this work.

After constructing a logical process from three freshly initialised processes, it is possible to compute the simple majority of the $b$-values from every component's black-box. Additionaly, after grouping three separate logical processes (lower-level) into another logical process (higher-level), it is also possible to keep track of $p$-value majorities. Correspondingly, the highest-level logical process consists of nine components (lowest-level) grouped into three logical processes. There will be three $b$-value majorities and one $p$-value majority. This hierarchic construction where logical constructs are embedded into one another is known as \emph{concatenation}, and has the advantage of polynomially lowering the failure probability of the resulting logical processes \cite{nielsen2010quantum}.

\subsection{Fault detectors}
\label{sec:fd}

The fault detectors used in the described distributed system are detecting faulty component processes. For each logical process there is a separate associated fault detector that interacts with the components. A fault detector consists of a set of low-level processes (called ancillae) which are initialised, used for communicating with the component processes and terminated. The output state of the ancillae is used to compute a syndrome: infer which component process is faulty. A fault detector contains also two variables: the Boolean \emph{faulty} indicates if the associated logical process is faulty or not, and the integer \emph{pos} points to the faulty component.

Fault detectors can check either $b$-values (the components are controls and the ancillae targets) or $p$-values (the other way around, the components are targets and the ancillae controls). Once more, without loss of generality,  the following fault detectors will be responsible only for $b$-values.

Ancillae are usual processes and can be affected by faults, which are required not to propagate to the components. At the same time, as process failures are probabilistic, a set of ancillae is used by the detector for reaching (with high probability) a trustful decision about the logical process (see Section~\ref{sec:compred}). A $b$-value detector responsible for a logical process protected against $b$-value faults has the process components as controls and the ancilla as target during the communication steps. If the ancilla holds a $p$-value set to heads, this will propagate to the components, but it would not influence the $b$-value protection. Again, if the ancilla holds a $b$-value fault, this will not propagate given the communication protocol. 

A logical process could be faulty beyond correction when a majority of the components is faulty. Assuming that all the communication between the logical processes is transversal (see Section~\ref{sec:trans}), and because process faults are uncorrelated, it would be improbable that such processes exist. Their existence would be a result of a high $\tau$ (called in the quantum computing literature the \emph{error threshold}), but can be mitigated by using more powerful encodings (e.g. surface code). Therefore, it is  further assumed that \emph{the right encoding  was chosen} for the logical process states, and that faulty components form a minority.

Computing a simple majority of correct processes from a set of components is equivalent to finding the faulty components forming a minority. For the distributed system example, a minority consists of at most one component. Two ancillae are required, the first one compares the $b$-values between the first and the second components, and the second ancilla the $b$-values between the second and the third components. 

The two ancillae are initialised in the same known state $q_a$. Let the component $b$-values be $b_0,b_1,b_2$; the ancilla output states after termination will be $q_{a1}=q_a+(b_{a1} \oplus b_0 \oplus b_1) + p_{a1}$, $q_{a2}=q_a+(b_{a2} \oplus b_1 \oplus b_2) + p_{a2}$. Considering that initially the ancillae are not faulty, the fault detector will extract two bits of information $s_1=b_0 \oplus b_1$ and $s_2=b_1 \oplus b_2$, indicating how the $b$-values compare pairwise between the components. The extension to faulty ancillae is presented in Section~\ref{sec:compred}.

The syndrome bits $s_1$ and $s_2$ encode the index of the faulty component process. For $s_1=s_2=0$ no faulty component exists, and the fault detector sets its \emph{faulty} flag to \emph{false}. For all the other syndrome values, the detector sets \emph{faulty=true}, and the faulty component index is computed by $pos=s_2*2 + s_1 - 1$.

\subsection{Fault corrector}

The fault corrector communicates with all the fault dectors in the system, and has a global overview of all the faults that were detected during the execution of the distributed computation. The global perspective has the following advantage: the corrector can observe if the modelled $\tau$ failure rates are valid or not; is the modelled failure rate to low?

In the presence of faults (signalled by the detectors), the fault corrector has two options: to either correct the faults, or to try and track their effect throught the distributed algorithm. The direct correction could introduce failures, and for this reason fault-tracking is more advantageous. Fault-tracking is performed based on commutativity properties: it is known how faults are transformed by both local and global operations. Hence, corrections are required only after the distributed computation was terminated and the output states were read out from the distributed system.

\subsection{Transversality}
\label{sec:trans}

The transversal application of a logical operation (local or distributed) is its decomposition into (local or distributed) operations applied on the component processes. For example, the logical local operation $G_l$ is the $n$-fold application of $G$ on each of the $n$ components.

Faults are propagated by inter-process communication. In this section, propagation is illustrated by a distributed system with two logical processes, each constructed from three component processes. Propagation will be mitigated by transversal communication.

Let $q_l^c$ be the logical state of the logical control, and $q_l^t$ the state of the logical target. It is further assumed that in both logical processes (control and target) at most one component has its $b$-value set to heads. Once more, it should be noted that the processes are not aware of their values. The logical states were transformed by transversal logical operations resulting in three equivalent component states in each logical process: $q_0^c=q_1^c=q_2^c$ and $q_0^t=q_1^t=q_2^t$. 

As the component states in both control and target are equivalent, transversal inter-process communication is implemented by forming pairs between control and target components. There are two possibilities: 1) the same control component is paired with each target component (see Figure~\ref{}); 2) each control component is paired with a different target component (see Figure~\ref{}). The second scenario corresponds to transversal inter-process communication.

For the first scenario, assuming that the component process $0_c$ (the component indexed $0$ in control) is used, pairs of the following form $(control, target)$ are built: $(0_c,0_t)$, $(0_c,1_c)$ and $(0_c,2_t)$. The three communication steps result in the updated $b$-values of the target components ($b_{i_t}=b_{i_t} \oplus b_{0_c}$). Assuming $b_{0_c}$ is heads, the control fault was propagated to the target components. At this point it cannot be guaranteed that the logical target is protected against a component's single heads $b$-value. Furthermore, the total cost of communication between the logical process is $c_1=3\times cost(heads)$.

The second scenario, the transversal inter-process communication, could result in the following three pairs being formed: $(0_c,0_t)$, $(1_c,1_t)$, $(2_c,2_t)$. Maintaining the assumption of the component $0_c$ being faulty in the $b$-value, after the three communication rounds only the state of $0_t$ would be negatively affected ($b_{0_t}=b_{0_t} \oplus b_{0_c}$). As a result, after executing this communication scenario, it can be guaranteed that the logical target is further protected against a single heads value of $b$. The total cost of communication between the logical process is $c_2 \leq 1\times cost(heads)$. Transversality minimises the communication cost between the logical processes, because $c_1 \geq c_2$.

\subsection{Computational redundancy}
\label{sec:compred}

Transversality is the key to constructing fault tolerant operations on logical processes, but is not applicable for maintaining a consistent set of component processes. A different technique has to be devised. In the presentation of the process model it was mentioned that faults can occur any time: the black-box coin is tossed at random time points. The toss could happen before each local or distributed operation. Faults are also allowed to occur before a process is terminated: after the last operation, but before returning its final state.

The solution is to \emph{continously check and correct} every logical process in the system. Checking is performed by the fault detectors and corrections are applied by the fault corrector. A detection round consists of multiple epochs, but a logical process is continously checked (multiple rounds). 

Section~\ref{sec:fd} introduced the fault detector and its use of ancillae, but ancillae were considered correct. In the presence of faults, the syndrome bits could be incorrect and trigger an unnecessary correction that would introduce more faults. The solution is to repeat during a detection round the syndrome extraction procedure multiple times, similar to a sequence of epochs (an example of computational redundancy). Every epoch requires a new pair of ancillae, and the fault detector will perform majority voting between the three pairs of extracted syndrome bits.

A freshly initialised process that was detected as being faulty could be either corrected or directly terminated and a new process instance would need to be initialised. During the execution of a process (between local and distributed operations), correction is the only option. Process operations could be delayed by the complete detection and correction procedures, because the detectors requires multiple executions in order to achieve a probabilistically consistent decision. Thus, fault-tolerance introduces significant resource and computational overheads.

\section{Conclusions}

\section{Acknowledgements}
SJD acknowledges support from the JSPS Grant-in-aid for Challenging Exploratory Research and 
JSPS KAKENHI Kiban B 25280034

\bibliographystyle{plain}
\bibliography{if}
\end{document}